\documentclass[10pt,twocolumn,aps,prb,showpacs,amsmath,amssymb,a4paper,superscriptaddress]{revtex4-1}

\usepackage{amsmath,amssymb,amsfonts,bm,graphicx,color,xparse,xstring,siunitx}
\usepackage[colorlinks=true,citecolor=black,linkcolor=black,urlcolor=blue,anchorcolor=black]{hyperref}

\usepackage[capitalise]{cleveref}  

\DeclareDocumentCommand\opr{m}{\ensuremath{\boldsymbol{#1}}} 

\def\ct{\dagger}
\def\degree{$^{\circ}$}

\def\spu{\uparrow}
\def\spd{\downarrow}

\definecolor{colspinup}{RGB}{215,48,39}
\definecolor{colspindown}{RGB}{69,117,180}

\begin{document}

\title{Robust band gap and half-metallicity in graphene with triangular perforations}

\author{S{\o}ren Schou Gregersen}
\email{sorgre@nanotech.dtu.dk}
\affiliation{Center for Nanostructured Graphene (CNG), DTU Nanotech, Department of Micro- and Nanotechnology,
Technical University of Denmark, DK-2800 Kongens Lyngby, Denmark}

\author{Stephen R. Power}
\email{spow@nanotech.dtu.dk}
\affiliation{Center for Nanostructured Graphene (CNG), DTU Nanotech, Department of Micro- and Nanotechnology,
Technical University of Denmark, DK-2800 Kongens Lyngby, Denmark}
\affiliation{Department of Physics and Nanotechnology, Aalborg University, Skjernvej 4A, DK-9220 Aalborg East, Denmark}

\author{Antti-Pekka Jauho}
\affiliation{Center for Nanostructured Graphene (CNG), DTU Nanotech, Department of Micro- and Nanotechnology,
Technical University of Denmark, DK-2800 Kongens Lyngby, Denmark}

\date{\today}


\begin{abstract}
Ideal graphene antidot lattices are predicted to show promising band gap behavior (i.e., $E_G\simeq 500$ meV) under carefully specified conditions.
However, for the structures studied so far this behavior is critically dependent on superlattice geometry and is not robust against experimentally realistic disorders.
Here we study a rectangular array of triangular antidots with zigzag edge geometries and show that their band gap behavior qualitatively differs from the standard behavior which is exhibited, e.g, by rectangular arrays of armchair-edged triangles.
In the spin unpolarized case, zigzag-edged antidots give rise to large band gaps compared to armchair-edged antidots, irrespective of the rules which govern the existence of gaps in armchair-edged antidot lattices.
In addition the zigzag-edged antidots appear more robust than armchair-edged antidots in the presence of geometrical disorder.
The inclusion of spin polarization within a mean-field Hubbard approach gives rise to a large overall magnetic moment at each antidot due to the sublattice imbalance imposed by the triangular geometry.
Half-metallic behavior arises from the formation of spin-split dispersive states near the Fermi energy, reducing the band gaps compared to the unpolarized case.
This behavior is also found to be robust in the presence of disorder.
Our results highlight the possibilities of using triangular perforations in graphene to open electronic band gaps in systems with experimentally realistic levels of disorder, and furthermore, of exploiting the strong spin dependence of the system for spintronic applications.
\end{abstract}

\maketitle

\section{Introduction} \label{sec:intro}

Two-dimensional materials continually gain interest and achieve huge advances towards industrial realization in a number of fields, particularly electronics and spintronics.
Graphene is the most studied material within the two-dimensional family~\cite{Novoselov2012} due to unique properties such as high electron mobilities~\cite{Mayorov2011} above \SI{e5}{cm.V^{-1}.s^{-1}}, gate-tunable carrier concentration~\cite{CastroNeto2009}, and predicted long spin-relaxation lengths~\cite{Han2014} of several \si{\um}.
These studies have led to substantial efforts in fabricating and processing clean graphene systems~\cite{Wang2013} as well as pushing the limits of nanostructuring e.g. by high-resolution lithography.~\cite{Oberhuber2013,Xu2013}
To realize graphene-based electronics and in particular transistors, opening a band-gap has been one of the main drivers of both theoretical and experimental work.
Many studies propose using structural modifications of graphene systems, such as nanoribbons~\cite{Han2007}, or superlattice structures imposed by periodic gating~\cite{Pedersen2012a,Low2011} or strain,~\cite{Pereira2009,Pereira2009a} to achieve a band gap.
More recent attempts have considered chemical modification through absorption, substitution, or sublattice symmetry breaking, for example, by doping.~\cite{Balog2010,Denis2010,Lherbier2013,Aktor2016}
Periodic patterning of graphene sheets, for example, periodic perforation to form so-called graphene antidot lattices (GAL) or nanomeshes, is of particular interest since theoretical predictions suggest the possibility of obtaining sizable band gaps.~\cite{Pedersen2008,Ouyang2011}
Several groups have realized these structures in the lab .~\cite{Eroms2009,Bai2010,Kim2010,Giesbers2012}
Band gaps induced in periodically patterned graphene are however very sensitive to disorder and defects.~\cite{Power2014}
Current fabrication methods will inevitably yield systems with a significant degree of disorder.
A clear experimental signature of minibands and -gaps has yet been elusive.
In the magnetic and spintronic areas, the possibility of making graphene magnetic or realizing graphene-based spintronics has also attracted a lot of attention.~\cite{Han2014}
It has been predicted that pristine graphene exhibits uniquely long spin-relaxations times~\cite{Han2014} \SI{\sim 1}{\us}, although to date experiments~\cite{Tombros2007,Han2010,Kamalakar2015} still find relaxation times at least two orders of magnitudes lower; reasons for this are still under debate.
Inducing magnetic ordering, or at least magnetic moments, is desirable in order to achieve tunable magnetism useful for magnetic information storage or spin-manipulation devices.
There have been many works, theoretical and experimental, studying magnetic moments induced by vacancy defects~\cite{Yazyev2007,Palacios2008,Nair2012,McCreary2012}, adatoms~\cite{Yazyev2007,McCreary2012,Power2011}, substrate coupling and molecular doping~\cite{Hong2012}.
Nanostructured graphene is also predicted to display significant spin polarization at certain extended edges, namely those with a zigzag (zz) geometry.~\cite{Son2006,Yu2008}
Recent experimental findings also support the prospect of magnetic zz-edges even with a reasonable amount of edge-roughness observed.~\cite{Tao2011, Hashimoto2014,Magda2014}

In this work, we propose using superlattices of triangular shaped GALs with entirely zz-edges to gain large spin polarization, as confirmed by previous \emph{ab initio} studies.~\cite{Furst2009,Zheng2009}
Graphene nanostructures which contain \emph{noncomplementary} zz-edges, e.g. triangles and Christmas-trees (stacked triangles), display unique global ferromagnetic order,~\cite{Yu2008,Furst2009,Zheng2009} as we also will illustrate for the GAL case in \cref{sec:res.pol} below.
In contrast, \emph{complementary} zz-edged nanostructures, e.g. zz-edged hexagons, rhombi (two triangles back to back), or straight nanoribbons, display antiferromagnetic ordering.~\cite{Son2006,Yu2008,Trolle2013}
Even before spin polarization is considered, we show through our tight-binding study how zz-edged triangular antidot lattices form exceptionally robust band gaps.
When the effects of spin are included, a similarly robust half-metallicity is displayed near the Fermi level, allowing for only either spin up or down states at a particular energy.
In contrast to the half-metallic behavior predicted for nanoribbon devices~\cite{Son2006}, triangular antidots naturally exhibit half-metallicity without the need for difficult side-gates and transverse electrical fields.
We envisage that triangular antidots could be fabricated, for example, through lithography using patterned hexagonal boron-nitride as a mask.
Hexagonal boron-nitride naturally etches into triangular-holes due to the different etch rates of the two species i.e. boron and nitrogen.~\cite{Jin2009}
Kinks or chirality within triangle edges may form during fabrication, but it is likely that they will still display a magnetic signature, albeit reduced, in accordance with theory for chiral graphene nanoribbons~\cite{Yazyev2011}.
Our findings suggest a realistic path towards fabricating realistic spin polarized graphene nanostructures which could act as components in graphene-based spintronic devices.

The remainder of this paper is organized as follows.
The system geometries and electronic and spin polarization models are described in \cref{sec:geom}.
Then we present our results in \cref{sec:res}, first considering several representative geometries in \cref{sec:res.unpol} of both zz-edged and armchair(ac)-edged triangles without spin polarization.
Next we focus on a single zz-edged antidot lattice and include spin-interaction in \cref{sec:res.pol}.
Finally we consider the robustness of our results by extending the tight-binding description in \cref{sec:res.3rdnn} and by considering the effect of positional disorder in \cref{sec:res.dis}.
In \cref{sec:concl}, we discuss our findings and other important considerations.

\section{Geometry and model} \label{sec:geom}
\begin{figure}
\centering
\includegraphics{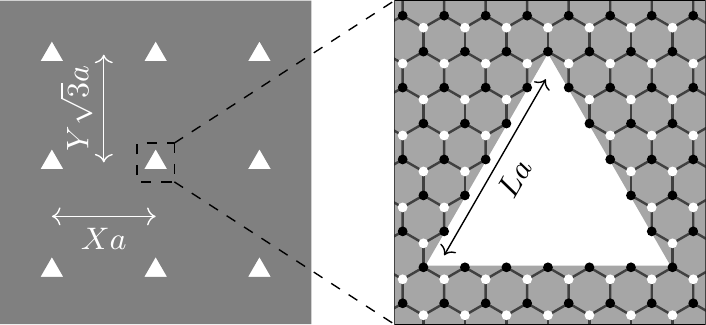}
\caption{
Schematic of the $\{25,15,5zz\}$ triangular antidot superlattice geometry (\emph{left}) and the approximately square unit cell with $X=25$, $Y=15$, and $L=5zz$-triangular antidot (\emph{right}).
The A and B sublattices of graphene are denoted by white and black circles respectively.
The antidot spacings are approximately 6~nm and the triangular side lengths are approximately 1~nm.
}
\label{fig:geom}
\end{figure}
Rectangular arrays of triangular antidots are considered as shown schematically in \cref{fig:geom}.
Specific geometries are denoted using $\{X,Y,L_{geo}\}$ where $X$ and $Y$ represent the inter-antidot spacings in the two in-plane directions, $L$ is the side length of the triangular antidot, and the index $geo = \mathrm{ac}$ or zz denotes the edge geometry of the triangles.
$X$ and $Y$ take integer values and the associated antidot separations are $X a$ and $Y \sqrt{3}a$ respectively, where the graphene lattice constant $a=\SI{2.46}{\angstrom}$.
The rectangular superlattice makes for an ideal testbed for antidot lattices.
The electronic properties change qualitatively with the superlattice dimensions, e.g. a semiconducting superlattice can become metallic and vice versa by changing the unit cell dimensions by just one lattice constant.~\cite{Pedersen2008, Petersen2011, Ouyang2011, Dvorak2013}
For any periodic external potential imposed onto graphene, for example an antidot lattice, if the Fourier transformed potential is zero at the Dirac points of pristine graphene a band gap \emph{cannot} form.
Antidot lattices for which the Fourier transformed potentials are \emph{nonzero} at the Dirac points have sizable band gaps.
This criterion is from hereon referred to as the \emph{periodicity selection rules}.~\cite{Dvorak2013}
For rectangular superlattices, due to the lattice orientation chosen, the periodicity selection rules depend critically on the $X$ spacing.
All antidot lattices for which $X=3p$ where $p=1, 2, 3, ...$ are semiconducting, while for all other antidot lattices the existence of gaps or not depends on the particular antidot.
Embedding the same triangular antidots into several rectangular superlattices which display different electronic behavior allows us to identify properties which arise due to the triangles themselves.
The triangular antidots we consider are aligned to have either zz-edges as shown in \cref{fig:geom} or ac-edges (not shown).
The latter ac-edged triangles are rotated by \ang{30} with respect to those in \cref{fig:geom} and the side length is scaled differently for the two orientations.
$L$ corresponds to a side length of $L a$ for the zigzag case and $(L \sqrt{3}a)$ for the armchair case.

Spin polarization at single-point defects, as well as that at zz-edges, is usually interpreted via Lieb's theorem.~\cite{Lieb1989}
The theorem states that the total ground state magnetic moment of a half-filled bipartite lattice is given by the sublattice imbalance, $M = \mu_B|N_A - N_B| \equiv \mu_B \Delta N $, where $N_A$ and $N_B$ are the are the number of sites belonging to each sublattice.
Creating a zz-triangle, such as that in \cref{fig:geom}, involves removing a different number of sites from the two sublattices and results in edge atoms belonging only to a single sublattice; with the orientation shown in \cref{fig:geom} this is sublattice B.
Accordingly, zz-triangles form \emph{nonzero} total magnetic moments, in full compliance with Lieb's theorem.~\cite{Furst2009, Zheng2009}
Rotating the antidot 180{\degree} flips the triangle orientation and also swaps the edge sublattice.
Thus the relative edge sublattices of two adjacent triangular antidots can be determined by a quick visual inspection.
The ac-triangle has both sublattices present at the edge and is not expected to exhibit spin polarization.~\cite{Yazyev2011}
We examine both the $\{X,Y,L\}=\{24,15,(5zz/3ac)\}$ and $\{25,15,(5zz/3ac)\}$ geometries; i.e. two geometries differing by $a$ along the $x$-direction and with either zz- or ac-edged triangular perforations; we later focus on the $\{25,15,5zz\}$ superlattice with the zz-triangle displayed in \cref{fig:geom}.
The side lengths of the zz- and ac-edged triangles are similar for these geometries.
The two triangle orientations highlight the fundamental differences between zz-edged triangular antidots, and the other antidot families represented by the ac-edged cases.

The calculations in \cref{sec:res.unpol} and \cref{sec:res.pol} are performed using a nearest neighbor (NN) tight-binding Hamiltonian
\begin{align}
H_{\sigma} &= \sum_{i} \epsilon_{i\sigma} {\opr c}_{i\sigma}^{\ct} {\opr c}_{i\sigma} + \sum_{ij} t_{ij} {\opr c}_{i\sigma}^{\ct} {\opr c}_{j\sigma} \;\, .
\label{eq:hamiltonian}
\end{align}
The operator ${\opr c}_{i\sigma}^{\ct}$ (${\opr c}_{i\sigma}$) creates (annihilates) an electron with spin $\sigma$ on site $i$ and the hopping parameter $t_{ij}$ takes the value $t=-2.7$eV when sites $i$ and $j$ are nearest neighbor sites  and is zero otherwise.
$|t|$ is taken as the unit of energy throughout the paper.
In \cref{sec:res.3rdnn} and \cref{sec:res.dis} we will consider an extension to a third nearest neighbor model (3NN) by including terms $t_2 = - 0.074|t|$ and $t_3 = -0.067|t|$ connecting second and third nearest neighbor sites respectively. \cite{Hancock2010}
The inclusion of $t_2$ results in a band-center shift, which we compensate for by adding a uniform on-site shift so that the Fermi energy lies at $E=0$.
Electron-electron interactions and the resulting spin polarization are included via spin-dependent on-site energy terms found from a self-consistent solution of the Hubbard model within the mean field approximation
\begin{align}
\epsilon_{i\sigma} &= \pm \frac{U}{2} m_i \;\, ,
\label{eq:hubbardU}
\end{align}
with $+$ for $\sigma=\spu$ and $-$ for $\sigma=\spd$.
, $m_i=\left<{\opr n}_{i\spu}\right> - \left<{\opr n}_{i\spd}\right>$ is the on-site magnetic moment, and ${\opr n}_{i\sigma}$ is the number operator.
We use the on-site Hubbard parameter $U = 1.33 |t|$ which has been shown to give results in good agreement with full \emph{ab initio} calculations for nanoribbon systems.~\cite{Yazyev2008,Yazyev2010}
The self-consistent Hubbard calculations are initiated with an antiferromagnetic guess, ${m_i=\pm c}$, with opposite signs used for the two sublattices A and B, and then iterated to convergence.

\section{Results} \label{sec:res}
\begin{figure}
	\centering
	\includegraphics[scale=0.9]{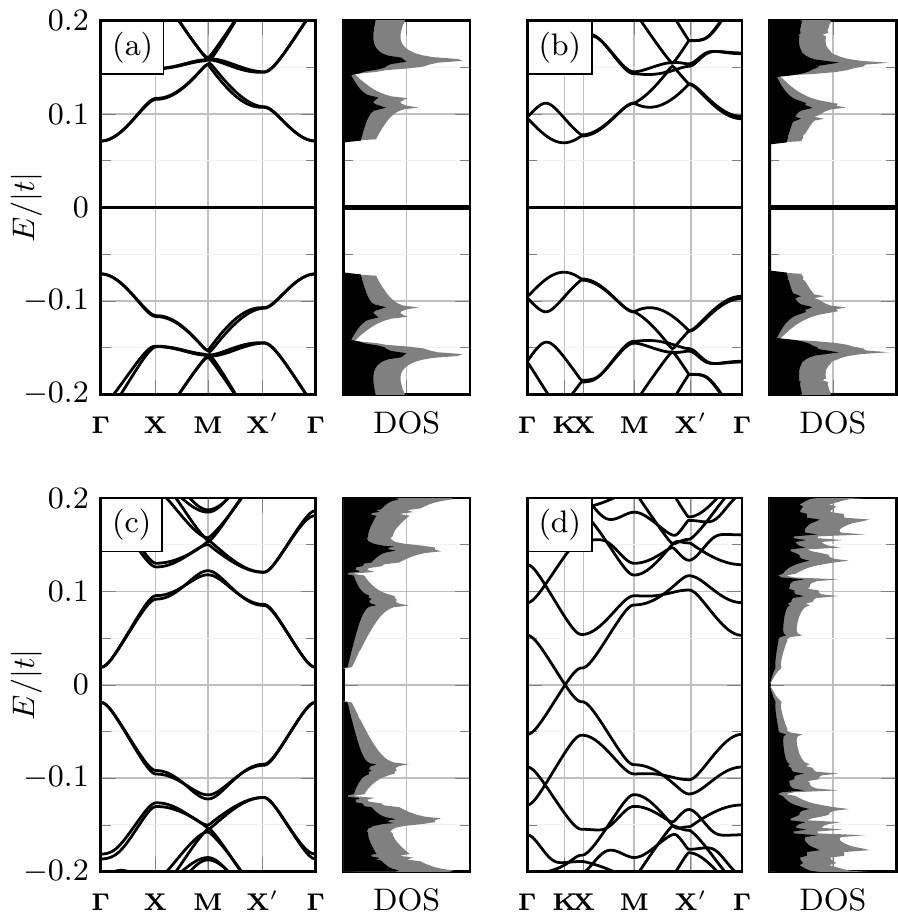}
	\caption{
		Unpolarized bandstructures and densities of states (DOS) of different triangular antidot systems.
		(a) $\{24,15,5zz\}$, (b) $\{25,15,5zz\} $, (c) $\{24,15,3ac\}$, and (d) $\{25,15,3ac\}$ geometries respectively.
		The DOS projected onto the edge sublattice B (\emph{black}) is shown together with the total DOS (\emph{gray}).
		The structures in (a) and (b) notably show very large and narrow peaks in the DOS at the Fermi level $E=0$.
	}
	\label{fig:bands}
\end{figure}

\subsection{Unpolarized antidots with different lattice geometries} \label{sec:res.unpol}
We first consider periodic structures of zz- or ac-triangular antidots in the $U = 0$ case.
The bandstructures of zz and ac-triangular antidots, together with their total density of states and that projected onto the (edge) B sublattice, are shown in \cref{fig:bands}.
The zz cases shown in \cref{fig:bands}a and \cref{fig:bands}b for the $\{24,15,5zz\}$ and $\{25,15,5zz\}$ geometries respectively, display both sizable band gaps and  dispersionless midgap states.
The 5-fold degenerate midgap states originate from the single-sublattice zz-edges.
The level of degeneracy is equal to the sublattice imbalance $\Delta N$, which also equals the number of zz-chains along the triangle edges $L = 5$.
Similar midgap states are also observed in other noncomplementary zz-edged nanostructures, e.g. triangular quantum dots~\cite{Jaskolski2011,Guclu2010,Sheng2012} and wide nanoribbons~\cite{Nakada1996,Hancock2010}, where the degeneracy is proportional to the global sublattice imbalance in the quantum dots, and to the local imbalance in the wide nanoribbons.
Such zz-edge states are localized on the edge sublattice.
Within the NN approximation states localized in a single sublattice remain completely dispersionless.
If higher order hopping parameters are included such states can also become dispersive, as we will discuss in \cref{sec:res.3rdnn} below.

The other characteristic of zz-edged triangular antidot lattices apparent from \cref{fig:bands}a and \cref{fig:bands}b is the formation of large electronic band gaps surrounding the dispersionless midgap states.
In comparison, the ac cases shown in \cref{fig:bands}c and \cref{fig:bands}d reveal that the $\{24,15,3ac\}$ is gapped and the $\{25,15,3ac\}$ is metallic.
These are in full compliance with periodicity selection rules, which in rectangular lattices predicts bands gaps only for cases with $X=3p$.
The zz-triangular antidots with large band gaps regardless of $X$ indicate a different band gap mechanism.
This hypothesis is supported by examining the band gaps of several triangular antidot lattices.
\begin{figure}
\centering
\includegraphics{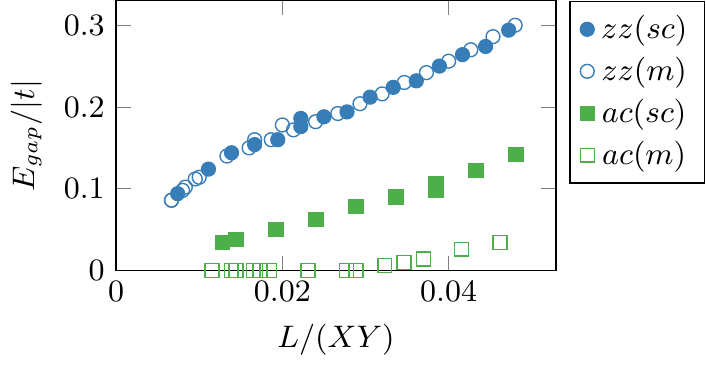}
\caption{
Unpolarized band gaps for various geometries as the dimensionless parameters $X$, $Y$, and $L$ are varied.
The zigzag (blue) and armchair (green) triangle geometries are divided into groups where the superlattice is expected to display semiconducting ($sc$, filled) or metallic ($m$, hollow) behavior, according to the periodicity selection rules.
For a rectangular superlattice this distinction depends solely on the value of $X$.
}
\label{fig:bandgaps}
\end{figure}

Pedersen \emph{et al.} demonstrated that a scaling behavior $E_{gap} \propto \frac{N_{rem}^{1/2}}{N_{tot}}\propto \frac{L}{XY}$ was followed by many gapped graphene antidot lattices\cite{Pedersen2008}, where $N_{rem}$ and $N_{tot}$ are, respectively, the number of atoms removed to form an antidot and the total number of atoms in the superlattice unit cell before the antidot atoms are removed.
In \cref{fig:bandgaps}, a linear behavior is clearly noted for those ac-edged systems with periodicity selection rules predicting semiconducting behavior (filled green squares) whereas those for metallic superlattices (hollow green squares) have zero band gap in almost all cases.
We associate the breakdown of this trend for metallic systems with large $\frac{L}{XY}$ to large antidots in small unit cells, where additional band gap behavior is now induced by small constrictions between the antidots.
The zz-triangles are meanwhile consistently gapped (blue circles), irrespective of the behavior predicted by periodicity arguments.
The band gap magnitude has an approximately linear dependence on $\frac{L}{XY}$, but the slope is much greater than the ac case.
The reason zz-edged triangular antidot lattices are consistently gapped is the global sublattice imbalance which induces \emph{sublattice symmetry breaking}.
Independent of the periodicity selection rules, sublattice symmetry breaking imposes an effective nonzero potential between sublattices in a similar manner to a \emph{mass} term, \textit{i.e.} a staggered on-site potential, with a different on-site potential for each sublattice.
In other systems where sublattice symmetry is broken, for example, by doping such a term also opens band gap.\cite{Lherbier2013,Aktor2016}
\begin{figure}
	\centering
	\includegraphics{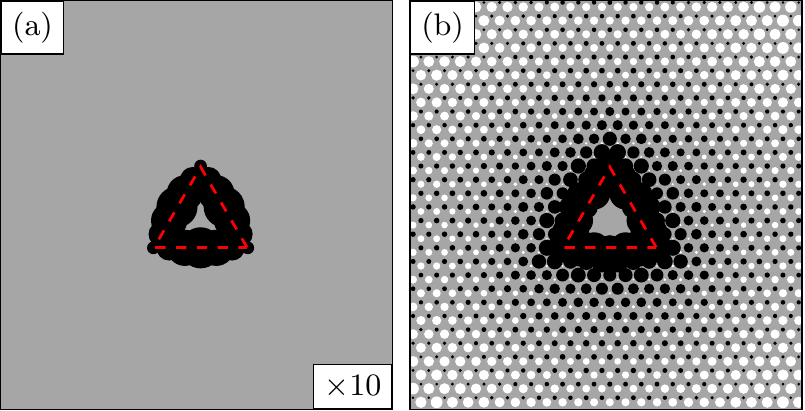}
	\caption{
		Unpolarized local density of states (LDOS) of the $\{25,15,5zz\}$ system.
		(a) The LDOS at the energy $E=0$ and (b) at the energy $E=0.1|t|$.
		A white (black) circle is placed on every site on the A (B) sublattice and its radius is scaled by the LDOS at that site.
		The zz-triangle edge is shown by a dashed red line.
		For clarity, the radii in (a) are reduced by a factor of $10$ relative to those in (b).
	}
	\label{fig:ldos}
\end{figure}

The sublattice-projected densities of states (DOS) for zz-triangle lattices in \cref{fig:bands}a and \cref{fig:bands}b show that each sublattice contributes equally to the DOS at all energies except at the $E=0$ edge states which reside only on the B sublattice.
However the local density of states (LDOS), shown in \cref{fig:ldos}, reveals a more complex picture.
The edge state localization is clear in \cref{fig:ldos}a where the LDOS is mapped at $E=0$ by circles whose radius is proportional to the LDOS at that site.
White and black circles are used for sites on the A and B sublattices respectively, and we note that only large black circles near the triangle edges are found at this energy.
Despite the equal contributions from sublattice projected DOS at other energies, the LDOS distributes inhomogeneously around the triangles.
This is shown in \cref{fig:ldos}b for the conduction band energy $E=0.1|t|$, where we note that the B sublattice contribution to the DOS is now spread throughout most of the unit cell, but is significantly larger near the triangle edges.
The A sublattice has a vanishing LDOS in this region and its DOS contribution is mostly distributed at sites midway between neighboring antidots.
The dispersion of the states at this energy is due to the large regions where both sublattices have a significant occupation.
The different electron distributions for the A and B sublattices suggest different effective scattering potentials for the different sublattices.
The inhomogeneous LDOS distribution, together with the band gap formation regardless of periodicity selection rules, suggests that sublattice symmetry breaking is the driving mechanism behind band gap formation and not the periodic selection rules usually forming band gaps in graphene antidot lattices.
Importantly, this suggests that band gap behavior in zz-triangles should be stable against geometrical variations as long as the sublattice imbalance is maintained.
Since the dimensions $X$ and $Y$ of the antidot lattice play a minor role, one may expect that lattices made of triangular zz-edged antidots are robust against disorder, as we discuss in \cref{sec:res.dis} below.

\subsection{Effect of spin polarization} \label{sec:res.pol}
A nonzero Hubbard-interaction ($U = 1.33 |t|$) leads to spin dependence in zz-edged triangle systems through the formation of magnetic moments $m_i$.
\begin{figure}
	\centering
	\includegraphics{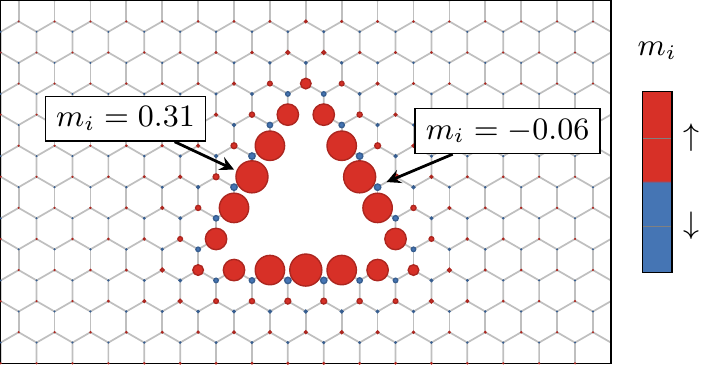}
	\caption{
	The magnetic moments surrounding a triangular antidot in the $\{25,15,5zz\}$ geometry.
	Spin up ($m_i > 0$, red) and spin down ($m_i < 0$,blue) moments are represented by circles whose
	radii are scaled by  $|m_i|$ at each site.
	The largest spin up ($m_i = 0.31$) and spin down ($m_i = -0.06$) moments are located respectively on an edge and on a site immediately next to the edge.
	The moments throughout the structure are antiferromagnetic i.e. the sign of a moment is determined by the sublattice on which it resides.}
\label{fig:moments}
\end{figure}
The self-consistent solution to the Hubbard model using the $\{25,15,5zz\}$ geometry is shown in \cref{fig:moments}.
Different superlattice and triangle dimensions always yield a similar pattern, namely a distribution with antiferromagnetic alignment between moments on the different sublattices.
The magnitude of the moments is maximum at the zz-edges, decreases slightly towards the corners of the triangles, and quickly decays perpendicular to the zz-edges.
Similar moment distributions have been reported in \emph{ab initio} studies of triangular perforations\cite{Furst2009}.
Triangles with large side lengths have long segments with approximately constant magnetic moments with a maximum $m_i \sim 0.31\mu_B$.
Only below $L < 5$ do these constant-moment segments vanish and the maximum moment decreases.
All the geometries considered are consistent with Lieb's theorem such that that $M = \sum_i m_i \mu_B =  \mu_B \Delta N \equiv  \mu_B L $.
The triangle corners are geometrically similar to the kinks arising in chiral graphene nanoribbons, which display a similar drop in moment values.~\cite{Yazyev2011}
The magnetic moment profile is found to be almost completely independent of the superlattice geometry, suggesting that nearby triangles do not influence each other unless they are very near.
\begin{figure}
\centering
\includegraphics{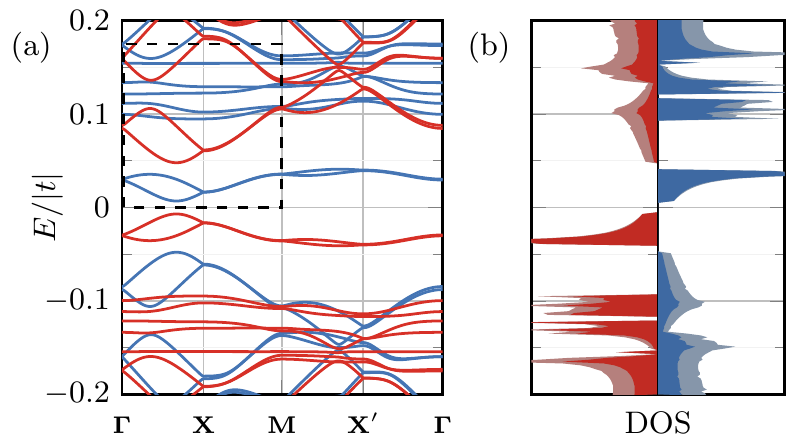}
\includegraphics{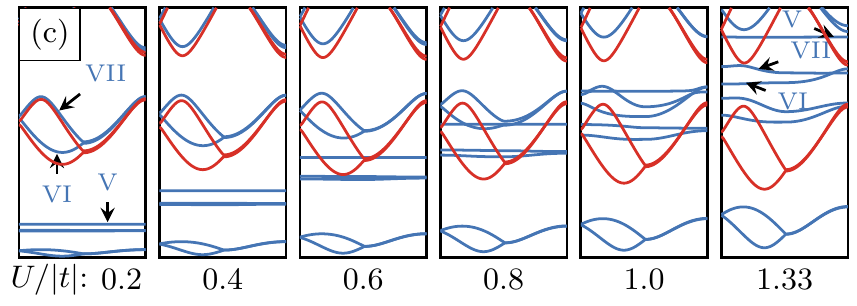}
\caption{
The effect of  Hubbard interaction $U$ on the $\{25,15,5zz\}$ geometry.
(a) The spin polarized band structures with spin up ($\spu$, \emph{red}) and spin down ($\spd$, \emph{blue}).
(b) The densities of states.  Projection on the edge sublattice B (\emph{darker} shading, $\spu$; \emph{red}, $\spd$: \emph{blue}), and the total DOS (\emph{lighter} shading, $\spu$: \emph{red}, $\spd$: \emph{blue})
(c) A zoom of the band structure (dashed box in (a)) for varying interaction strength $U/t=0.2.\cdots1.33$.
}
\label{fig:bands.pol}
\end{figure}

The spin-split band structure of the $\{25,15,5zz\}$  system is shown in \cref{fig:bands.pol}a, together with the spin up (red) and spin down (blue) DOS in \cref{fig:bands.pol}b.
As before, the lighter regions show the total DOS and the darker regions show its projection onto the edge B sublattice.
There are a number of key differences from the unpolarized band structure of the same geometry system in \cref{fig:bands}b compared to the spin polarized bandstructure in \cref{fig:bands.pol}a.
The five-fold degenerate dispersionless bands are no longer present at zero energy and the band gap is considerably reduced by the presence of dispersive bands at the energies $E=\pm 0.02|t|$.
These bands have opposite spin orientations on either side of $E=0$, as do the five low-dispersive non-degenerate bands in the energy range $(\pm) 0.1|t| \rightarrow 0.15|t|$.
To examine the formation of this bandstructure the Hubbard-interaction $U$ is varied from a low $U=0.2|t|$ to $U=1.33|t|$ in \cref{fig:bands.pol}c, left to right.
The band structures in these panels correspond to the region shown by the dashed box in \cref{fig:bands.pol}a.
We denote three low energy spin down bands V, VI, and VII at low Hubbard-interaction strength $U=0.2|t|$, the fifth through seventh lowest energy spin down bands in this region.
In the unpolarized bandstructure, band V corresponds to one of the five-fold degenerate dispersionless bands whereas VI and VII form the conduction bands.
The V, VI, and VII bands are labeled both at the left- and right-most panels for clarity.
These panels reveal how the formerly degenerate and dispersionless bands undergo different degrees of spin splitting.
The highest of these (V) initially at low $U$ (left) appears below both bands VI and VII and finally at high $U$ (right) appears above said bands.
The degree of spin splitting is determined by the LDOS distribution and the magnitude of the magnetic moments.
High degrees of spin splitting can be attributed to a LDOS localized around areas with large magnetic moments, which is confirmed by examining the spin polarized LDOS.

\begin{figure}
	\centering
	\includegraphics{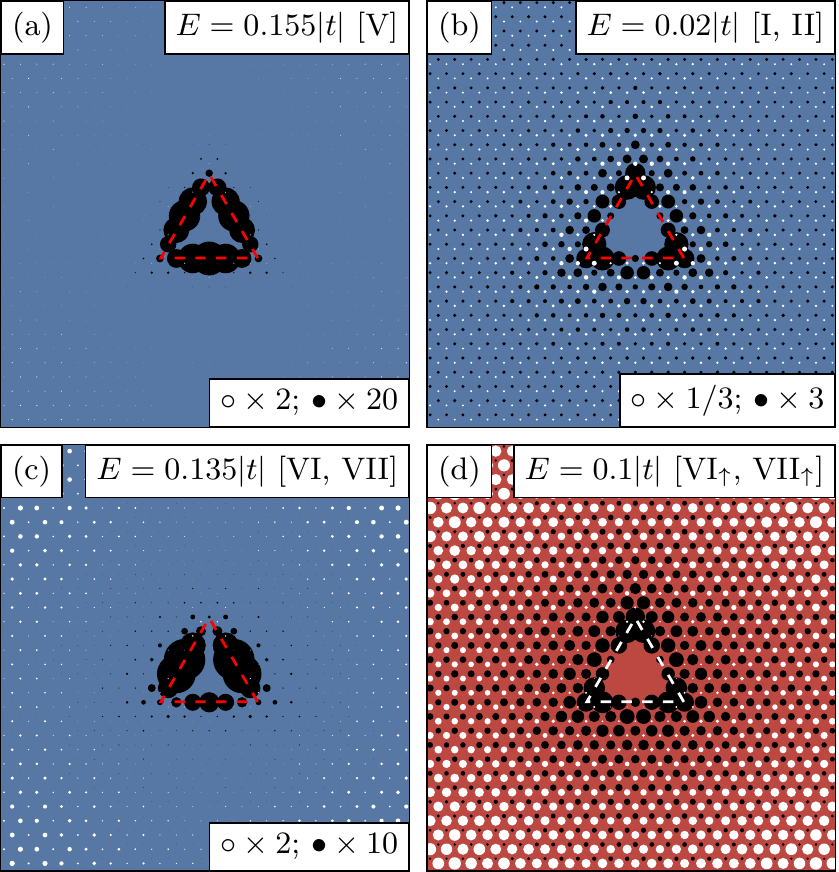}
	\caption{
		Polarized LDOS of $\{25,15,5zz\}$.
		(a) At energy $E=0.155|t|$, (b) $E=0.02|t|$, (c) $E=0.135|t|$, and (d) $E=0.1|t|$.
		A white (black) circle is placed on every site on the A (B) sublattice and its radius is scaled by the LDOS at that site.
		The zz-triangle edge is shown by a dashed line.
		For clarity, the radii in (a,b,c) are reduced by factors denoted in the \emph{lower right corner} relative to those in (d).
	}
	\label{fig:ldos.pol}
\end{figure}
At $U=1.33|t|$ and $E=0.155|t|$, corresponding to the zz-edge states band V and shown in \cref{fig:ldos.pol}a, the LDOS is localized almost entirely on magnetic edge-sites, consistent with a large degree of spin splitting.
Meanwhile, the LDOS of the spin polarized conduction bands at $E=0.02|t|$ shown in \cref{fig:ldos.pol}b is mostly localized near the triangle corners which have smaller magnetic moments, consistent with a small degree of spin splitting.
Further, the dispersion of the conduction bands is shown to emerge due to a non-zero occupation of the A sublattice as shown in \cref{fig:ldos.pol}b.
In the unpolarized case bands VI and VII define the conduction band edge, but as $U$ increases (see \cref{fig:bands.pol}c), the spin-down versions flatten and increase in energy, whereas the spin up versions broaden and decrease slightly in energy.
We noted earlier that the unpolarized cases displayed LDOS contributions from both sublattices, which overlapped to form dispersive conduction bands.
When spin polarized, this distribution is quite different for each spin.
The LDOS of the spin down band shown in \cref{fig:ldos.pol}c is localized almost entirely on B sublattice sites near the center of the zz-edge sections, which leads to a flattening of the dispersion and an upwards energy shift.
Conversely, the LDOS of the spin up bands shown in \cref{fig:ldos.pol}d is localized both on the B sublattice near the antidot corners and on sites from both sublattices further away from the triangle.
The more homogeneous distribution of the spin up bands leads to further broadening and a weaker downwards shift from spin-splitting.

\begin{figure}
\centering
\includegraphics{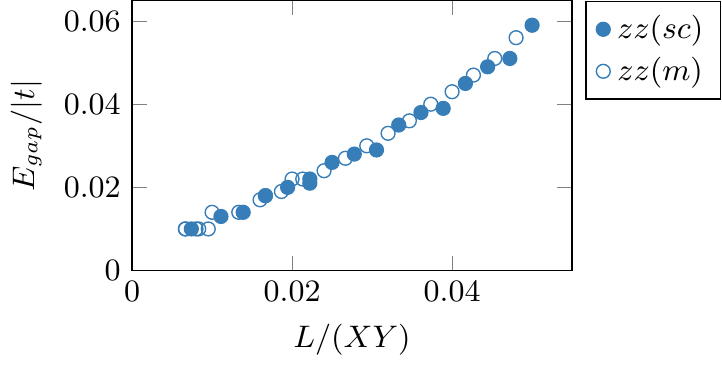}
\caption{
Polarized band gaps for various geometries with zz-triangular antidots embedded varying $X$, $Y$, and $L$.
The geometries are divided into groups where the superlattice is expected to display semiconducting ($sc$) or metallic ($m$) behavior, which for a rectangular superlattice depends solely on the value of $X$.
}
\label{fig:bandgaps.pol}
\end{figure}
The band gaps for spin polarized zz-triangles are shown for a range of geometries in \cref{fig:bandgaps.pol}, where we note a decrease of approximately one order of magnitude compared to the unpolarized cases.
In fact, the gaps of semiconducting ac-triangles are larger than those for polarized zz-triangles.
However, spin polarized zz-edged antidots display another interesting feature.
The dispersive states surrounding the band gap are completely spin polarized, so that a spin-selective half-metallicity can be induced by small $E_F$ shifts applied using a back gate.
This suggests that such geometries may be employed in a range of spintronic components to filter spins of different orientations.

Many of the features we have described in both unpolarized and polarized zz-triangles depend on the inhomogeneous electron distributions and in particular the localization on the edge sublattice and near zz-edges.
It is important to determine if such features are artifacts of the NN model we employ for our calculations, and whether they are robust in the face of disorder.
The latter point is of interest as many effects induced by superlattices tend to vanish at any realistic disorder.~\cite{Power2014}
We now briefly address both issues.

\begin{figure}
	\centering
	\includegraphics{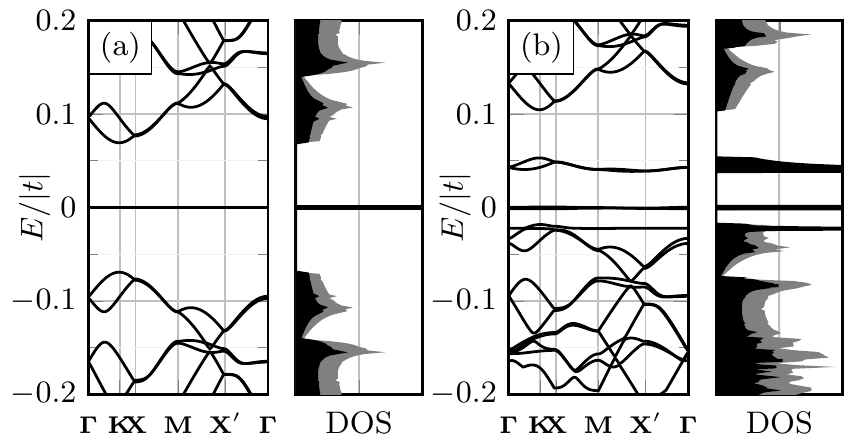}
	\includegraphics{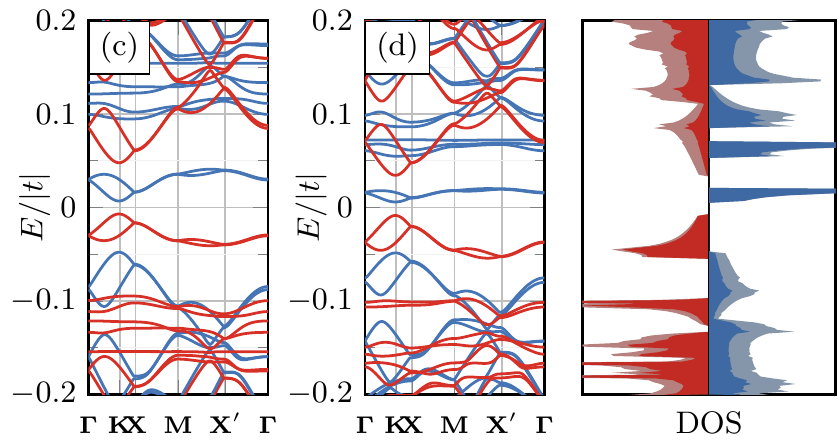}
	\caption{
		Band structures for the $\{25,15,5zz\}$ geometry within first (NN) and third (3NN) nearest-neighbor tight-binding models.
		For NN and 3NN spin unpolarized as well as the 3NN spin polarized band structures, the DOS is also shown.
		(a) NN and (b) 3NN without spin polarization, (c) NN and (d) 3NN with spin polarization.
		(a) and (c) are reproduced from respectively \cref{fig:bands}b and \cref{fig:bands.pol}a.
		The DOS projected onto the edge sublattice B (\emph{darker} shading, $\spu$: \emph{red}, $\spd$: \emph{blue}) is shown together with the total DOS (\emph{lighter} shading, $\spu$: \emph{red}, $\spd$: \emph{blue}).
	}
	\label{fig:bands.3rd-nn}
\end{figure}

\subsection{Effect of higher order hopping terms}\label{sec:res.3rdnn}
Within the NN model, states which occupy only a single sublattice appear completely dispersionless.
In comparison, a 3NN model enables intra-sublattice coupling by the inclusion of the 2NN terms, and the parametrization we use has been shown to accurately describe zz-nanoribbons.~\cite{Hancock2010}
For the unpolarized case, we note that the introduction of these additional hopping terms leads to an energy splitting of the previously degenerate midgap states, see \cref{fig:bands.3rd-nn}a and \cref{fig:bands.3rd-nn}b.
This leads to a shift of the Fermi energy relative to the bulk valence and conduction bands in order to satisfy half-filling, increasing the electron-hole asymmetry already introduced by the 2NN hoppings.
The NN-model band gap can be identified in the 3NN band structure between the energies $E=-0.025|t|$ and $E=0.1|t|$, but is now slightly smaller and more importantly contains multiple midgap states.
In particular a dispersive channel opens at $E \sim 0.05|t|$, similar to that seen near zz-ribbon edges when a 3NN model is employed.\cite{Chang2012}
Disregarding these midgap states, the 3NN band gap between $E=-0.025|t|$ and $E=0.1|t|$ scales similarly to the NN model when varying the system dimensions.
The emergence of dispersive states in the band gap could of course limit the applicability of these systems.
However we note that in many cases they have either very little dispersion, or are spaced far enough apart in energy, so as to still offer reasonable band gap or transport gap behaviors.

Considering the polarized case, the band structures and DOS in \cref{fig:bands.3rd-nn}c and \cref{fig:bands.3rd-nn}d are remarkably similar despite the large changes we have discussed in their associated unpolarized versions.
The most significant change now between NN and 3NN models is the expected (minor) electron-hole asymmetry.
Notably the system remains half metallic with spin-dependent dispersive states close to the Fermi level.
The excellent agreement between NN and 3NN models in this case can be understood by the fact that the features introduced by the additional 3NN terms in the unpolarized case, namely dispersion and splitting of the midgap states, also result independently from the inclusion of the spin-dependent potentials.
We note that the 3NN model, both with and without spin polarization, also agrees qualitatively with previous \emph{ab initio} calculations, which display similar band structures.~\cite{Furst2009}
Although the 3NN model serves to correct the missing inter-sublattice interaction, it appears that the most important behavior in polarized systems is captured by the lower order NN model.

\subsection{Robustness against disorder}\label{sec:res.dis}
\begin{figure}
\centering
\includegraphics{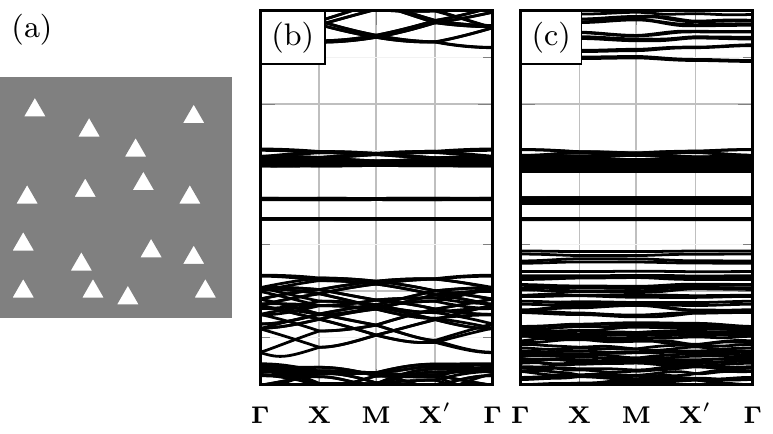}
\includegraphics{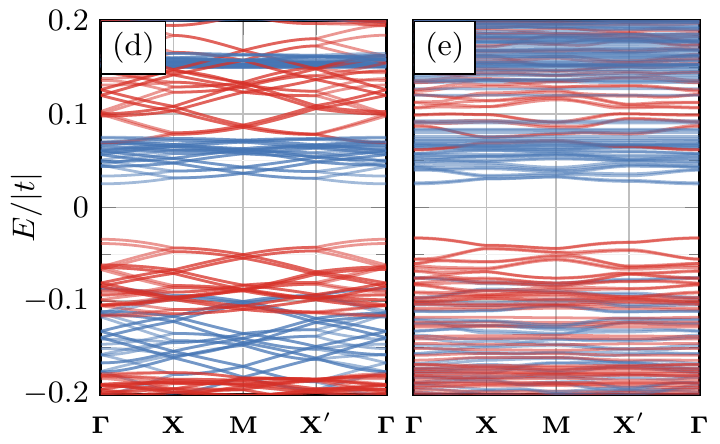}
\caption{
(a) Schematic of a disordered 4-by-4 array of $\{15,9,5zz\}$ triangular antidots. (b) Pristine band structure.
(c) Disordered system for $U=0$ (d) Pristine system for $U=1.33|t|$, (e) Disordered system for $U=1.33|t|$.
}
\label{fig:dis.zz}
\end{figure}
\begin{figure}
\centering
\includegraphics{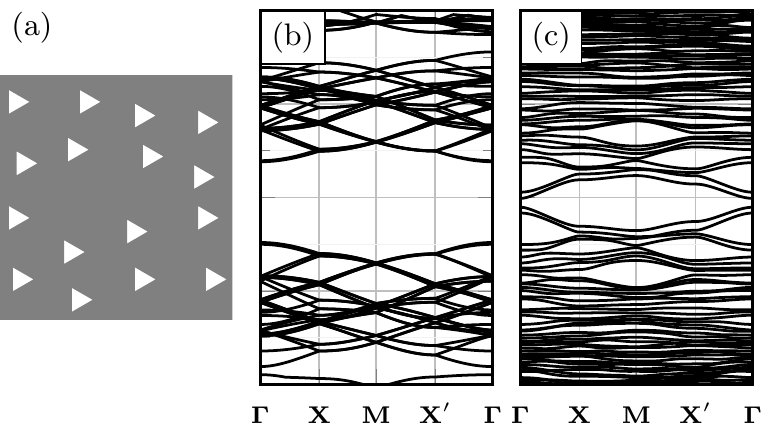}
\caption{
(a) Schematic of a disordered 4-by-4 array of $\{15,9,3ac\}$ triangular antidots.
(b) Band structure for a pristine system, $U=0$ (c) Band structure for a disordered system, $U=0$.
}
\label{fig:dis.ac}
\end{figure}

One of the major obstacles in inducing band gaps using graphene antidots is that the large gaps predicted in atomically precise systems are extremely fragile in the presence of even mild geometrical disorders.~\cite{Power2014}
The gap mechanism for usual antidot arrays, namely the periodicity selection rules, relies on pristine conditions and regular antidot spacing.
We have shown that zz-edged triangular antidots behave very differently from other antidots, and that their behavior arises from the breaking of sublattice symmetries around individual antidots.
We further demonstrated that these effects were independent of the superlattice geometry, which suggests also that they should be more stable than, for example, ac-edged antidots, in the face of disorder.
While a full-fledged disorder analysis is beyond the scope of the present paper, we highlight the essential effects by examining a $4\times4$ geometrically disordered array of antidots in a repeated superlattice.
We present one particular random configuration but also note that an additional 10 different configurations have been examined all showing qualitatively the same behavior.
The triangle centers are randomly shifted by $\Delta \opr{r} = \{\delta x a, \delta y (\sqrt{3} a)\}$ with $\delta x = \delta y \leq 3$, as shown for zz-triangles in \cref{fig:dis.zz}a and ac-triangles in \cref{fig:dis.ac}a.
The same size triangles as before are considered, but for computational efficiency we use smaller $\{X=15,Y=9\}$ ``blocks" to compose the supercell, essentially cutting down on the amount of pristine graphene between perforations.
Note that according to the periodicity selection rules these superlattice geometries are predicted to form band gaps.\cite{Dvorak2013}
Pristine band structures calculated within the $4\times4$ framework are shown for unpolarized and polarized $\{15,9,5zz\}$ systems in \cref{fig:dis.zz}b and \cref{fig:dis.zz}d respectively, and for the (unpolarized) $\{15,9, 3ac\}$ system in \cref{fig:dis.ac}b.
All calculations here were performed within the 3NN model.
We note that larger gaps are present in all cases due to the reduced $X$ and $Y$ values, and that significant folding of the bands has occurred due to the larger supercell.
However the same qualitative behavior for zz-edges from \cref{fig:bands.3rd-nn}b and ac-edges from \cref{fig:bands}c is evident.
The gapped region in the unpolarized zz-edged antidot case, \cref{fig:dis.zz}c, is partially quenched due to a small energy spreading of states.
The polarized bands, \cref{fig:dis.zz}e, show even less variance relative to the ordered case.
In contrast, similar levels of disorder quench the gap almost entirely for ac-edged triangles, as demonstrated in \cref{fig:dis.ac}c, consistent with results for other disordered antidot systems whose band gap emerges from periodicity selection rules.~\cite{Power2014}
Despite the same level of geometrical disorder, zz-edged triangles appear far more robust compared to ac-edged triangles.
In comparison, the spin polarized band structure of the $\{25,15,5zz\}$ geometry, which display smaller band gaps in \cref{fig:bands.3rd-nn}c and \cref{fig:bands.3rd-nn}d, might in the presence disorder significantly reduce the spin polarized band gap.
Nevertheless, even with reduced band gaps in the spin polarized case we expect the band gap of the unpolarized bands and the half-metallicity of the polarized bands to remain at these levels of disorder.
Two additional types of disorder could have a significantly larger effect: orientation angle disorder and edge disorder.
The former has the effect of dividing the triangles into smaller regions of zz-edges connected by kinks.
Reducing the length of the zz-edged regions will in turn reduce the sublattice symmetry breaking and the band gap formed in the superlattice.
The latter type of disorder has the same effect of reducing the length of the zz-edged regions, but additionally can introduced localized scatterers which could induce additional states with in the previous band gap, severely reducing the final band gap of such a superlattice.
What is truly different for zz-edged triangular antidots compared to, for example ac-edged antidots, is that while intra-antidot disorder like angle and edge disorder might quench the band gap of both shapes, inter-antidot disorder will have a much larger effect on the ac-edged antidots.

\section{Conclusion} \label{sec:concl}

We have discussed the electronic properties of triangular antidots systems in graphene sheets, with a particular focus on zigzag edged geometries whose geometry breaks the symmetry between graphene's two sublattices.
In order to shed light on the possibility of magnetic states at such edges, we have analyzed systems in both the spin polarized and the unpolarized cases.
We further have illustrated the robustness against disorder by individually displacing the antidots of a 4-by-4 array unit cell.
Spin unpolarized superlattices of triangular zz-edged antidots form band gaps significantly larger than similarly sized ac-edged counterparts.
Gap-opening occurs irrespective of conventional rules governing the formation of band gaps in, for example, ac-edged triangular antidots and scales with the triangular antidot side length.
Furthermore zz- as opposed to ac-edged triangles are far more robust against geometric disorders.
We conclude that these unique features are caused by a gap-opening mechanism related to sublattice-symmetry breaking.
In contrast to conventional graphene antidot lattices, this mechanism is less sensitive to experimentally unavoidable imperfections in lattice spacings.
The zz-edged triangular antidots become half-metallic over a wide range of energies when spin polarization is included, with a high degree of spin selectivity achievable by gating.
Spin splitting of the unpolarized band structure leads to the emergence of dispersive spin-dependent states and subsequent reduction of the band gaps compared to the unpolarized cases.
The half-metallic behavior of zz-edged triangles also appears more robust against geometric disorder compared to ac-edged counterparts.
These findings suggest a robust path to realize devices based on nanostructured graphene with robust band gaps.
Further, devices with half-metallic and spin-selective properties appear feasible.

\section{Acknowledgments} \label{sec:ackno}
The Center for Nanostructured Graphene (CNG) is sponsored by the Danish Research Foundation, Project DNRF103.

\bibliography{references}

\end{document}